\documentclass[pra,reprint,doublecolumn,superscriptaddress, nobalancelastpage]{revtex4-1}

\usepackage{tikz}

\usepackage{etoolbox}
\patchcmd{\section}
  {\centering}
  {\raggedright}
  {}
  {}
\patchcmd{\subsection}
  {\centering}
  {\raggedright}
  {}
  {}
\usepackage{upgreek}
\usepackage{tcolorbox}
\usepackage[sort&compress]{natbib}
\usepackage{soul}
\usepackage{graphicx}
\usepackage{epstopdf}
\usepackage{verbatim}
\usepackage{float}
\usepackage{sidecap}
\sidecaptionvpos{figure}{c}
\usepackage{lipsum}
\usepackage{setspace}
\usepackage{dcolumn}
\usepackage{bm}
\usepackage{amssymb}
\usepackage{amsmath}	
\usepackage{mathtools}
\usepackage{color}
\usepackage{multirow}
\usepackage{lipsum}
\usepackage[a4paper]{geometry}
\newgeometry{top=1.25cm,right=1cm,left=1.5cm,bottom=1.25cm}
\usepackage[colorlinks=true, urlcolor=blue, pdfborder={0 0 0}]{hyperref}
\hypersetup{
linkcolor=blue,
citecolor=blue
}

\draft

\newcommand{\kdotp}{$\textbf{k}\cdot\textbf{p}$ }
\newcommand{\spss}{$sp^{3}s^{*}$ }
\newcommand{\GaBiAs}{GaBi$_{x}$As$_{1-x}$ }
\newcommand{\GaBiAsGaAs}{GaBi$_{x}$As$_{1-x}$/GaAs }

\begin{document}

\setcounter{secnumdepth}{0}

\title{\textbf{ \large{Towards low-loss telecom-wavelength photonic devices by designing GaBi$_{x}$As$_{1-x}$/GaAs core$-$shell nanowires}}}

\author{Muhammad Usman} \email{musman@unimelb.edu.au} \affiliation{School of Physics, The University of Melbourne, Parkville, 3010, Victoria, Australia.}
\maketitle
\onecolumngrid

\noindent
\textcolor{black}{
\normalsize{\textbf{Nanowires are versatile nanostructures, which allow an exquisite control over bandgap energies and charge carrier dynamics making them highly attractive as building blocks for a broad range of photonic devices. For optimal solutions concerning device performance and cost, a crucial element is the selection of a suitable material system which could enable a large wavelength tunability, strong light interaction and simple integration with the mainstream silicon technologies. The emerging \GaBiAs alloys offer such promising features and may lead to a new era of technologies. Here, we apply million-atom atomistic simulations to design GaBi$_{x}$As$_{1-x}$/GaAs core$-$shell nanowires suitable for low-loss telecom-wavelength photonic devices. The effects of internal strain, Bi Composition ($x$), random alloy configuration, and core-to-shell diameter ratio ($\rm \rho_D$) are analysed and delineated by systematically varying these attributes and studying their impact on the absorption wavelength and charge carrier confinement. The complex interplay between $x$ and $\rm \rho_D$ results in two distinct pathways to accomplish 1.55 $\upmu$m optical transitions: either fabricate nanowires with $\rm \rho_D \geq$ 0.8 and $x \sim$15\%, or increase $x$ to $\sim$30\% with $\rm \rho_D \leq$ 0.4. Upon further analysis of the electron hole wave functions, inhomogeneous broadening and optical transition strengths, the nanowires with $\rm \rho_D \leq$ 0.4 are unveiled to render favourable properties for the design of photonic devices. Another important outcome of our study is to demonstrate the possibility of modulating the strain character from a compressive to a tensile regime by simply engineering the thickness of the core region. The availability of such a straightforward knob for strain manipulation without requiring any external stressor component or Bi composition engineering would be desirable for devices involving polarisation-sensitive light interactions. The presented results document novel characteristics of the GaBi$_{x}$As$_{1-x}$/GaAs nanowires with the possibility of myriad applications in nanoelectronic and nanophotonic technologies.}}}
\\ \\
\twocolumngrid

Semiconductor nanowires made up of III-V alloys are a topic of great interest for fundamental research as they form a promising system with highly tunable electronic and optical characteristics through engineering of their geometry parameters and composition~\cite{Dasgupta_advm_2014, Zhuang_Advm_2011}. In terms of applications, they are an excellent absorber and emitter of light, and provide viable pathways for carrier collection and transport. The advances in the fabrication techniques for nanowires have opened possibilities for growth on silicon substrates, which offers rich opportunities for novel integrated nanoeletronic and nanophotonic technologies including photodetectors, waveguides, light-emitting diodes (LEDs), lasers, solar cells and light-sensors~\cite{Yao_NNano_2013, Tomioka_Nature_2012, Chen_NatureP_2011, Wang_Nanoletters_2003, Yan_Nature_Photo_2009, Eaton_Nature_Review_2016}. Furthermore, there have been cross-disciplinary proposals to design nanowires for novel applications such as cell imaging and solar to fuel conversion~\cite{Yang_Nanoletters_2010}. Consequently, there is an immense research interest in designing semiconductor nanowires with tailored characteristics suitable for the next generation technologies.   

Nanowires have been shown to epitaxially grow without forming dislocations based on a large number of III-V lattice-mismatched materials and there has been significant progress in terms of monolithic integration with a variety of substrates such as silicon~\cite{Tomioka_Nature_2012, Dimakis_NanoLett_2014} or graphene~\cite{Munshi_NanoLett_2012}. A highly promising feature of the nanowire geometry is the prospect to assemble structures in the form of core$-$shell, where the core and shell regions could be engineered by their relative size and by selecting materials with very different lattice constants~\cite{Royo_JPDAP_2017}. This allows tremendous flexibility to custom design nanowires with properties suitable for a wide range of applications, making them highly versatile nanostructures. The core$-$shell nanowires involving conventional III-V materials such as InGaAs, AlGaAs, and InGaP have been around for a number of years, and their electronic and optical properties have been extensively studied in the literature~\cite{Kim_Nanoletters_2006, Ren_Adv_Mat_2014, Ma_Nano_Lett_2014, Dai_NanoLett_2014, Dimakis_NanoLett_2014}. However, the fabrication of nanowires formed by the highly mismatched alloys such as \GaBiAs is currently at a primitive stage and the preliminary experimental studies have been recently reported, which investigated their structural, strain and optoelectronic properties~\cite{Ishikawa_Nanoletters_2015, Oliva_arXiv_2019, Matsuda_JAP_2019}. Up until now, there is no theoretical study in the literature which described the optoelectronic properties of the \GaBiAsGaAs nanowires. There are a number of open questions concerning the introduction of Bi in the \GaBiAsGaAs nanowires and its impact on strain, electronic wave functions, and inter-band optical absorption strengths. A theoretical investigation based on the comprehensive and systematic set of atomistic simulations such as presented in this work could provide a timely guidance to the future experiments and may significantly contribute in the advancement of this exciting and emerging field of research. 

The \GaBiAs materials, which are formed by dilute concentrations of Bi atoms in GaAs, offer an exquisite control over the bandgap wavelengths. It has been shown that approximately 60-90 meV reduction in the band gap energy can be achieved per 1\% increase in Bi incorporation in the GaAs material~\cite{Janotti_PRB_2002, Zhang_PRB_2005, Usman_PRB_2011}, which makes these materials highly attractive as building blocks for a variety of light absorption or emission devices working in a broad range of wavelengths encompassing telecommunication to mid and far infra-red spectral range (1 to 10 $\upmu$m)~\cite{Bismuth_containing_compounds_2013, Broderick_bismide_chapter_2017, Fluegel_PRL_2006}. A recent study based on spectroscopic ellipsometry has shown that the refractive index for \GaBiAs increases with Bi fraction and is slightly larger than that for GaAs material~\cite{Tumenas_PSSc_2012}. The most promising property of the \GaBiAs material system is related to the large spin-orbit coupling associated with the GaBi material which makes it possible to tune the spin-orbit splitting energy above the band gap energy at telecom-wavelength (1.55 $\upmu$m) for the \GaBiAs alloys at relatively low Bi concentrations ($\approx$ 13\%)~\cite{Batool_JAP_2012, Usman_PRB_2011}. This characteristic, not accessible in the conventional III-V materials, is unique for the bismide alloys and is expected to drastically suppress the non-radiative Auger recombination and the inter-valence band absorption (IVBA) processes in the bismide-based photonic devices, which otherwise plague the efficiency of the devices made up of the conventional semiconductor materials~\cite{Phillips_IEEEJSTQE_1999, Sweeney_PSSB_1999, Broderick_SST_2012}. Therefore, the design of the \GaBiAsGaAs core$-$shell nanowires is an important emerging area of research with tremendous potential for the implementation of photonic devices with low internal losses and improved temperature stability~\cite{Ishikawa_Nanoletters_2015, Oliva_arXiv_2019, Matsuda_JAP_2019}.

To enable the design of optoelectronic devices based on the semiconductor nanowires with optimised characteristics, it is of paramount importance to thoroughly understand the relationship between the physical attributes of a nanowire, such as its geometry and material composition, and the resulting electronic structure and optical transition strengths. An in-depth analysis of the nanowire properties can also provide a useful guidance for the future experiments to target optimised physical parameters during the fabrication stage. Recognising the importance of such theoretical study, this work has employed million-atom atomistic simulations to investigate the strain, electronic and optical properties of the \GaBiAsGaAs core-shell nanowires. The valence force field and tight-binding methods used to compute strain and electronic structure have been carefully parametrised in accordance with the published density functional theory (DFT) based study of the GaBi material~\cite{Janotti_PRB_2002, Usman_PRB_2011}. Subsequently, the electronic structure calculations based on our methods demonstrated an excellent agreement with the experimental measurements performed on bulk GaBiAs alloys~\cite{Usman_PRB_2011}, strained GaBiAs alloys~\cite{Usman_PRB_2013}, and the GaBiAs/GaAs quantum well structures~\cite{Usman_PRA_2018}. We also note that several published studies based on the DFT theory~\cite{Kudrawiec_JAP_2014, Polak_SST_2015, Bannow_PRB_2016} and experimental measurements~\cite{Donmez_SST_2015, Balanta_JoL_2017, Zhang_JAP_2018, Dybala_APL_2017, Collar_AIPA_2017, Usman_PRB_2011, Usman_PRB_2013, Broderick_PRB_2014} have also shown a good agreement with our theoretical predictions, further affirming the high-level accuracy of our computational techniques. Consequently, the results reported in this work not only provide a highly reliable fundamental understanding of the \GaBiAsGaAs nanowire properties, but also highlight a set of optimal parameters for the design of low-loss telecom-wavelength photonic devices, which can be expected to provide a clear direction for the future experiments on the fabrication of the bismide nanowires. 

\begin{figure*}[t]
\includegraphics[scale=0.4]{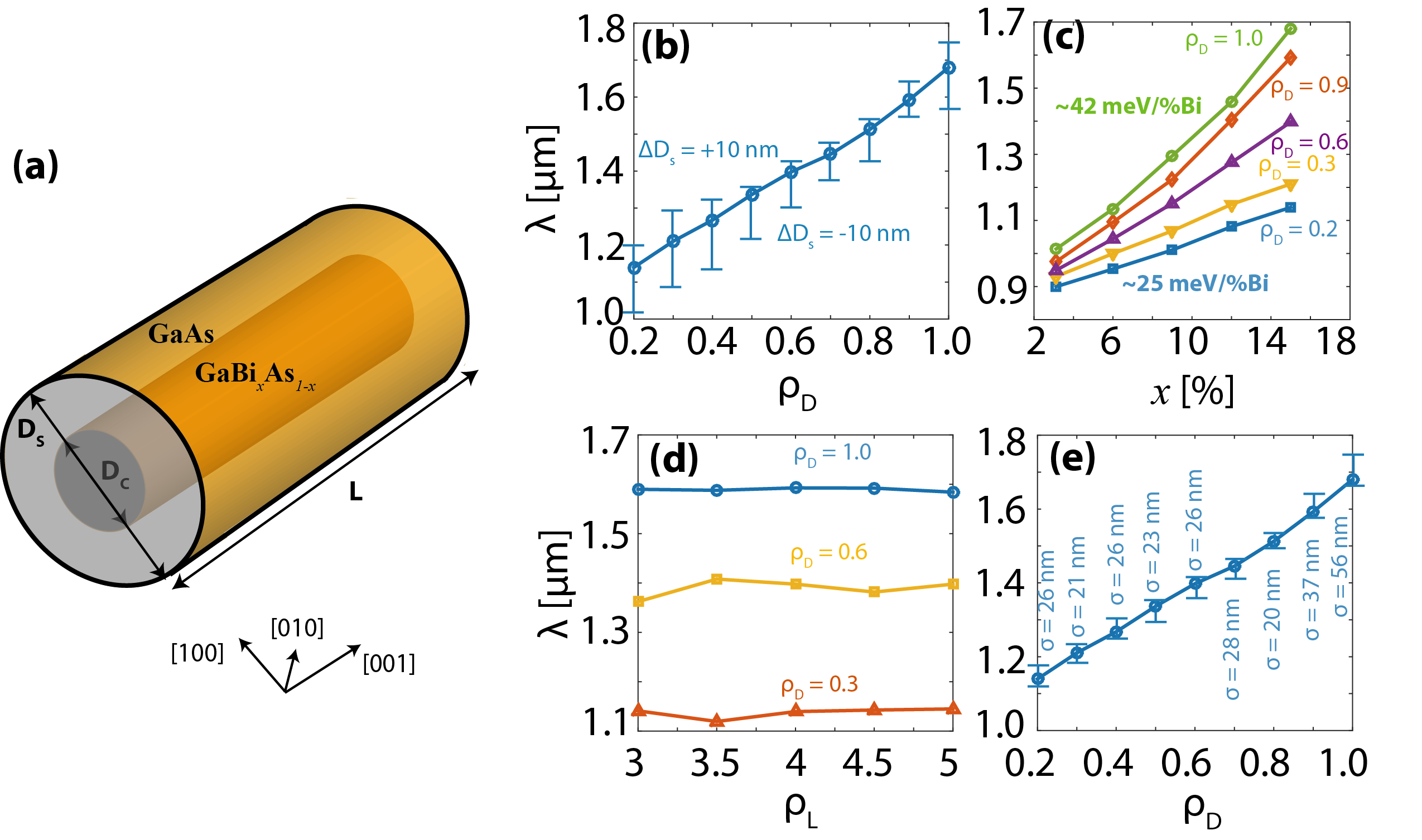}
\caption{\textbf{Wavelength engineering.} \textbf{(a)} Schematic diagram of the investigated \GaBiAsGaAs core$-$shell nanowire is illustrated. The nanowire consists of a GaAs shell with diameter $\rm D_S$, a \GaBiAs core with diameter $\rm D_C$, and a length of L along the [001] direction. The ratio between the core and shell diameters is defined as $\rm \rho_D$ = $\rm \frac{D_C}{D_S}$, and a ratio of the length to the shell diameter is denoted by $\rm \rho_L$ = $\rm \frac{L}{D_S}$. We note that the shell diameter of nanowire is same as the overall diameter of the nanowire structure. \textbf{(b)} The ground-state inter-band absorption wavelength ($\rm \lambda$) is plotted as a function of $\rm \rho_D$ for $\rm D_S$=20 nm, L=80 nm, and $x$=15\% . The variation bars at each data point indicate a change in $\rm \lambda$ with respect to a $\rm \pm$10 nm change in $\rm D_S$.  \textbf{(c)} The plots of $\rm \lambda$ as a function of Bi fraction ($x$) in the core region for various values of $\rm \rho_D$ at L=80 nm and $\rm D_S$=20 nm. \textbf{(d)} The plots of $\rm \lambda$ as a function of $\rm \rho_L$ for different values of $\rm \rho_D$ at $\rm D_S$=20 nm and $x$=15\%. \textbf{(e)} The plots of $\rm \lambda$ as a function of $\rm \rho_D$ at $\rm D_S$=20 nm, L=80 nm, and $x$=15\%, where the variation bars exhibit a change in $\rm \lambda$ with respect to five different random alloy configurations for \GaBiAs in the core region.}
\label{fig:Fig1}
\end{figure*} 

This work is based on a systematic set of simulations in which the physical parameters of the \GaBiAsGaAs core-shell nanowires such as the relative diameters of the core and shell regions, the Bi composition of the core region, and the length of the nanowire are varied and their impact on the resulting optoelectronic character is delineated. The \GaBiAs material has shown to exhibit a strong alloy disorder effect on its electronic properties~\cite{Gogineni_APL_2013, Imhof_APL_2010, Usman_APL_2014, Usman_PRA_2018}, which cannot be captured by the traditional effective-mass or \kdotp modelling techniques~\cite{Alberi_PRB_2007, Broderick_SST_2013}. Here, we have applied atomistic simulations to quantitatively capture the impact of the random placements of the Bi atoms in the \GaBiAs alloy formation and provide an accurate understanding of the resultant inhomogeneous broadening of the bandgap wavelength. The presented results uncover an interesting interplay between the Bi composition of the nanowire and core-to-shell diameter ratios. We predict that the optical absorption at 1.55 $\upmu$m can be achieved by either fabricating a thick core region with around 15\% Bi compositions, or by increasing the Bi fraction to 30\% while fabricating narrow diameters of the core region. As the spin-orbit splitting energies are expected to be below the bandgap energies at Bi compositions above about 13\%~\cite{Usman_PRB_2011}, both of these nanowire compositions would meet this condition and therefore expected to lead to suppressed internal loss processes. Next, we analyse these two fabrication scenarios by comparing the optical transition strengths and propose that high Bi compositions with narrow core diameters are optimised nanowire designs to achieve a stronger inter-band optical transition strength and a relatively weaker inhomogeneous broadening of the ground state optical transition at 1.55 $\upmu$m wavelength. This provides a favourable direction for the future fabrication experiments on \GaBiAsGaAs core$-$shell nanowires. Our results also reveal that the strain arising due to the lattice mismatch between the GaAs and \GaBiAs materials demonstrates a large shift in its character from a compressive to tensile regime when the core-to-shell diameter ratio is increased towards 1. This is expected to open new avenues for the fabrication of strain engineered nanowires, with the possible application in devices requiring polarisation-sensitive light interactions (absorption or emission)~\cite{Usman_PRB2_2011}.    

\section{Results and Discussions} 

In this work, we have simulated \GaBiAsGaAs core$-$shell nanowires based on atomistic simulations, where the simulation domain for the largest structure consisted of about 3.5 million atoms. The large-scale simulations are crucial to attain a comprehensive understanding of the strain, electronic and optical character of the investigated nanowires. A brief summary of the computational methods is provided at the end of the paper and a detailed account of the methodologies can be found in the \textcolor{blue}{supplementary information} document. The \GaBiAsGaAs core-shell nanowires are constructed in Zincblende structure, which is commensurate with the recent fabrication reports~\cite{Ishikawa_Nanoletters_2015}. The strain profiles are computed by first relaxing the nanowire geometry in accordance with the atomistic valence force field (VFF) method~\cite{Keating_PR_1966}. 

The parameters of interest associated with the confined electronic states such as electron and hole energies and the corresponding wave functions are computed by solving an sp$^3$s$^*$ tight-binding Hamiltonian at the $\Gamma$ point of the Brillouin zone. In contrast to the simplified models widely used in the literature such as valence-band anticrossing (VBA) and \kdotp methods~\cite{Alberi_PRB_2007, Broderick_SST_2013}, the presented atomic-resolution description of the \GaBiAs alloy allows an understanding and incorporation of the atomic-scale effects, including random alloy configurations (random positioning of the Bi atoms forming different size and type of clusters)~\cite{Usman_PRB_2011} and roughness at the GaAs and \GaBiAs interface~\cite{Singh_JAP_1985}, which could strongly impact the inhomogeneous broadening and transport properties of an alloyed nanowire and play an important role in the performance of the nanowire based devices. Accordingly, our results document a reliable and quantitative estimate of the nanowires characteristics. One of the key aims of this study is to provide a direct guidance to the experimental groups working on the implementation of nanowire based photonic devices. In this regard, we have specifically focused on the design of nanowire parameters which are generally controllable during the fabrication stage such as core/shell diameters and Bi composition. Based on a thorough investigation, we have suggested a nanowire design which leads to a strong ground state transition strength and a weak inhomogeneous broadening at 1.55 $\upmu$m ground state transition wavelength, the two desired attributes for the future photonic technologies. 

\noindent
\textbf{\textit{\textcolor{blue}{Wavelength tuning.}}} Figure~\ref{fig:Fig1} (a) illustrates the schematic diagram of the investigated core-shell nanowire structure with a cylindrical symmetry around the [001]-axis. The core region is made up of the \GaBiAs material with a diameter D$\rm _C$ and Bi fraction $x$. The shell region is the GaAs material with a diameter D$\rm _S$. The length of the nanowire along the [001] direction is L. In this work, we investigate nanowire properties as a function of the relative diameter of the core region with respect to the shell diameter and the length of the nanowire. Therefore, we define two parameters of interest as $\rm \rho_D$=$\rm \frac{D_C}{D_S}$ and $\rm \rho_L$ = $\rm \frac{L}{D_S}$. In the remainder of the paper, we will primarily focus on the understanding of the strain, electronic and optical properties of nanowires with reference to $\rm \rho_D$ and $\rm \rho_L$. The Bi atoms are placed in the core region by randomly replacing As atoms in accordance with the intended Bi concentrations ($x$). We consider five different random configurations of Bi atoms at each $x$ and all of the presented results are computed by averaging over the five alloy configurations. The large size of the \GaBiAs core region (consisting of 50000 to 1.5 million atoms) along with the investigated five random alloy configurations are expected to provide an accurate description of the alloy randomness and the associated inhomogeneous broadening in consistent with the published literature~\cite{Usman_PRM_2018}. 

Figure~\ref{fig:Fig1} (b) plots the ground state optical transition wavelength ($\lambda$=1.24/(E$_1$-H$_1$) $\upmu$m) as a function of the nanowire diameter ratio $\rm \rho_D$. Here, E$_1$ and H$_1$ are the energies of the lowest and the highest confined electron and hole states, respectively. The Bi fraction in the core region is fixed at 15\%. The solid curve is based on the shell diameter D$\rm _S$=20 nm and the core diameter is varied as D$\rm _C$=$\rm \rho_D$D$\rm _S$. Our results indicate that the wavelength is red shifted when the core diameter is increased, keeping the shell diameter fixed. A value of $\rm \rho_D$ close to 0.9 results in the desired 1.55 $\upmu$m optical transition wavelength. The error bars indicate a variation of $\pm$10 nm in the shell diameter at each value of $\rho_D$. The plotted results exhibit that such variation in the core and shell diameters at a fixed ratio $\rm \rho_D$ translates to about 0.2 $\upmu$m variation in $\lambda$. 

Figure~\ref{fig:Fig1} (c) shows the plot of $\lambda$ as a function of the Bi fraction $x$ in the core region of nanowire for various values of $\rho_D$ at D$_S$=20 nm. The wavelength is red shifted when $x$ is increased, irrespective of $\rm \rho_D$. However, the red shift per \% Bi is larger when the core diameter is closer to the shell diameter ($\rm \rho_D$ closer to 1). Our calculations show a red shift of $\sim$25 meV/\%Bi and $\sim$42 meV/\%Bi at $\rm \rho_D$=0.2 and 1.0, respectively. Importantly, from these computed results, we infer that the fabrication of nanowires with large $\rho_D$ will require significantly less Bi fraction in the core region to target 1.55 $\upmu$m wavelength. For example, at $\rho_D$=0.9, about 15\% Bi is required compared to about 30\% for $\rho_D$=0.2. Furthermore, these results highlight two pathways to achieve the desired 1.55 $\upmu$m wavelength for the telecommunication photonic devices: either by increasing the Bi fraction at small values of $\rho_D$ or by increasing $\rho_D$ with relatively small Bi fractions. These two options can be explored to optimise other properties such as optical absorption strengths as will be done in the later part of the paper. Here, we also note that the ground state transition at 1.55 $\upmu$m wavelength require relatively large Bi fraction in heterostructure nanowires ($\geq$15\%), compared to the bulk strained \GaBiAs material where the tight-binding theory and the photo-reflectance measurements indicated that only about 10\% Bi fraction lead to a similar red shift in $\lambda$~\cite{Usman_PRB_2013}. 

In Figures~\ref{fig:Fig1} (b) and (c), we have kept the length of the nanowires (L) fixed at 80 nm and only changed the diameter ratio ($\rho_D$) and Bi fraction ($x$). The length of the nanowire is also controllable during the fabrication process and hence is a parameter of interest. Figure~\ref{fig:Fig1} (d) plots the inter-band transition wavelength as a function of the ratio $\rm \rho_L$ by changing L from 60 nm to 100 nm for the three selected values of $\rho_D$. From these results, we predict that the ground state transition wavelength is only weakly dependent on $\rm \rho_L$. Consequently, we infer that the analysis and the understanding developed in this work will be applicable for the larger lengths of the nanowires provided that the overall ratios $\rm \rho_D$ and $\rm \rho_L$ are consistent with the investigated range. 

\begin{figure*}[t]
\includegraphics[scale=0.27]{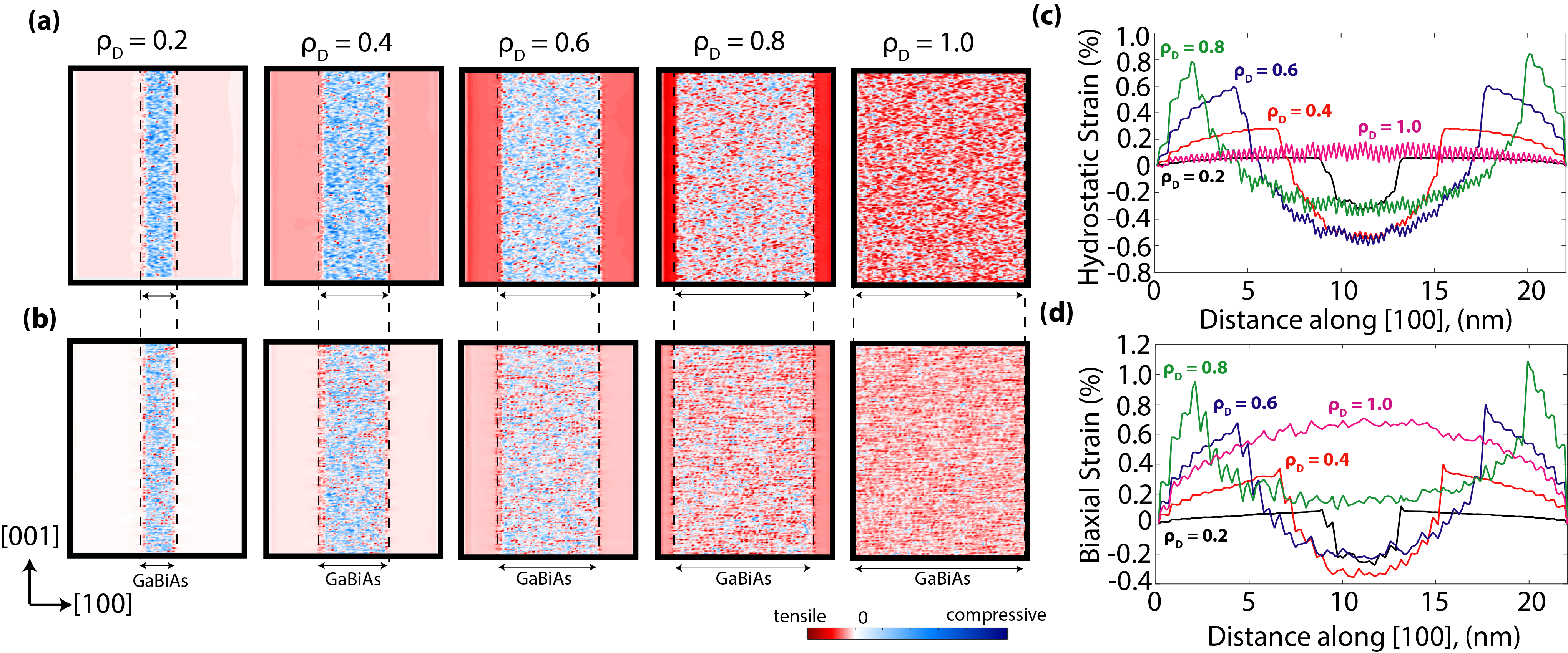}
\caption{\textbf{Strain analysis.} \textbf{(a)} The plots of the hydrostatic strain component ($\epsilon_H = \epsilon_{xx} + \epsilon_{yy} + \epsilon_{zz} $) are shown for the \GaBiAsGaAs nanowires with $x$=15\%, $\rm \rho_L$=4, a few selected values of $\rm \rho_D$. The color plots show the strain profile in a 2D [010]-plane through the center of the nanowire. The blue (red) color region indicates the presence of a compressive (tensile) strain. \textbf{(b)} The same as \textbf{(a)} but for the biaxial strain component ($\epsilon_B = 2\epsilon_{zz} - \epsilon_{xx} - \epsilon_{yy} $). \textbf{(c)} The 1D plots of the hydrostatic strain component are shown through the center of the nanowire with $x$=15\%, $\rm \rho_L$=4, a few selected values of $\rm \rho_D$. The computed strain at each data point along the [100] direction represents an average over all data points in the corresponding [100] plane and for five different random configurations of Bi atoms. \textbf{(d)} The same as \textbf{(c)} but for the biaxial strain component.}
\label{fig:Fig2}
\end{figure*}

The \GaBiAs material belongs to a special class of III-V alloys also known as highly-mismatched semiconductors. This is due to the fact that heavy Bi is much larger than As and is also less electronegative, therefore its addition to the GaAs material leads to very different characteristics compared to the conventional III-V alloys~\cite{Bismuth_containing_compounds_2013}. More specifically, the Bi atoms largely retain their electronic character when replacing the As atoms in the formation of the \GaBiAs alloys, and as a result, the impact of Bi on the electronic structure of the GaAs material does not follow the conventional virtual crystal approximation; rather it is dictated by an interaction of the Bi related resonant states with the valence band edges of the GaAs material via a band anti-crossing interaction (BAC)~\cite{Broderick_SST_2012, Usman_PRB_2011}. An important consequence of this character is that when the Bi atoms form pairs or clusters, their interaction with the GaAs electronic structure vary over a considerably large range depending upon the type and size of the Bi cluster~\cite{Usman_PRB_2011}. This leads to an associated large broadening of the inter-band optical absorption wavelengths, known as the inhomogeneous broadening of the ground state transition. For the \GaBiAs quantum well heterostructures, an inhomogeneous broadening of 23-33 meV has been reported in the literature~\cite{Usman_APL_2014, Broderick_SST_2015}. As the inhomogeneous broadening of the optical transition wavelength plays an important role in the performance of optoelectronic devices~\cite{Dery_IEEEJQE_2005}, here we investigate its strength for the nanowire structures. For this purpose, we randomly vary Bi atom positions in the core region while keeping the overall Bi concentration fixed at 15\%. Figure~\ref{fig:Fig1} (e) plots the transition wavelengths as a function of $\rho_D$, where the error bars now indicate the inhomogeneous broadening of the wavelengths ($\sigma$) based on the five different random configurations of the Bi atoms. Interestingly, we find that the inhomogeneous broadening of the wavelengths for nanowires is similar to the previously reported values for the \GaBiAs quantum well structures. Our results also indicate that the broadening becomes stronger when the diameter of the core region is closer to the shell diameter ($\rm \rho_D \geq$ 0.9). This is somewhat counter intuitive, given that the actual number of Bi atoms per unit volume of the nanowire core region is smaller ($\approx$0.01675 per nm$^3$) for the nanowire with $\rm \rho_D$=1.0 compared to the $\approx$0.02 per nm$^3$ for the $\rm \rho_D$=0.2 nanowire, implying that relatively lesser number of Bi pairs/clusters should exist in the core region for the larger diameters. However, as it will be clear in the next section, the smaller width of the shell region for the $\rm \rho_D \geq$ 0.9 nanowires implies that the compressive strain is drastically relaxed in the core region and therefore the effect of the alloy randomness becomes stronger, as would be expected for the bulk \GaBiAs material compared to the compressively strained \GaBiAs~\cite{Usman_PRB_2013}. 

Based on the discussion above and the data plotted in Figure~\ref{fig:Fig1}, it is summarised that the technologically relevant 1550 nm wavelength absorption can be realised in \GaBiAsGaAs core$-$shell nanowires either by engineering $\rm \rho_D$ or by tuning the Bi fraction in the core region at a fixed geometry configuration. Notably, Figure~\ref{fig:Fig1}(c) shows that the 1550 nm wavelength is possible by increasing the Bi fraction, independent of both $\rm \rho_D$ and $\rm \rho_L$. Later, it will be further shown that this pathway provides a clear advantage to maximise the absorption strength. 

\noindent
\textbf{\textit{\textcolor{blue}{Strain analysis.}}} In the case of the lattice-mismatched semiconductor heterostructures, the internal strain plays an important role in the electronic and optical character by directly impacting the electron and hole confinement energies, the spatial distributions of the wave functions, and the polarisation direction of the light absorption or emission~\cite{Usman_PRB2_2011, Usman_IEEE_2009}. Typically, when a larger lattice constant material such as \GaBiAs is surrounded by a smaller lattice constant material such as GaAs (for example in a quantum well, quantum dot, or a nanowire), the larger lattice constant material is expected to be under a compressive strain. However, the actual strain profile in the heterostructure could be a complex function of a number of parameters, such as the relative dimensions of the two materials, the type of heterostructure such as quantum well, dot or wire, the shape of heterostructure such as planar or axial, and the magnitude of the mismatch between the lattice constants. In our study, the net strain in the nanowire region will be a function of both the ratio of core-to-shell diameters $\rm \rho_D$ and the Bi fraction $x$ of the core region. Therefore, it is important to investigate the interplay between the two parameters to understand the net strain profile and its impact on the optoelectronic character of the nanowires. 

The strain is simulated by performing the atomistic valence force field relaxation of the nanowire structure over the large sizes of supercells consisting of up to 3.5 million atoms. The diagonal ($\epsilon_{xx}, \epsilon_{yy}, \epsilon_{zz}$) and shear ($\epsilon_{xy}, \epsilon_{yz}, \epsilon_{zx}$) components of the strain tensor are computed from the relaxed atomic positions~\cite{Lazarenkova_APL_2004}. We also calculated the hydrostatic strain ($\epsilon_H = \epsilon_{xx} + \epsilon_{yy} + \epsilon_{zz} $) and biaxial strain ($\epsilon_B = 2\epsilon_{zz} - \epsilon_{xx} - \epsilon_{yy} $) components, which are typically investigated to understand the strain effects on the electronic and optical properties of nanostructures~\cite{Usman_IEEE_2009}.

Figure~\ref{fig:Fig2} (a,b) plots the hydrostatic and biaxial components of the computed strain for a few selected nanowire geometries represented by $\rm \rho_D$ at $x$=15\%. The corresponding in-plane ($\epsilon_{xx}$) and vertical (($\epsilon_{zz}$)) strain components are plotted in the \textcolor{blue}{Supplementary Information Figure S1}. These plots exhibit strain profiles along a two-dimensional [001]-plane through the center of the nanowire region. Overall, we observe a similar trend for both strain components inside the \GaBiAs core region: the strain is highly compressive at small values of $\rm \rho_D$ and becomes increasingly tensile for $\rm \rho_D \geq$ 0.8,  where a negative (positive) sign of a strain component denotes its compressive (tensile) character. We also notice that for $\rm \rho_D$ = 0.6 and 0.8, the GaAs shell region is significantly tensile strained. This is expected as a much larger \GaBiAs core region exerts tensile pressure on the narrow GaAs shell region and leads to the stretching of the bonds in that region. 

To provide a quantitative estimate of the strain profiles, we also plot one-dimensional cuts as a function of the distance along the [100] direction through the center of the nanowires, which are shown in Figure~\ref{fig:Fig2} (c,d). These plots exhibit an interesting character of the strain with respect to $\rm \rho_D$. At $\rm \rho_D$ = 0.2, both hydrostatic and biaxial strains are highly compressive inside the core region and nearly zero in the shell region. This character is typically expected for a heterostructure consisting of a large lattice constant material (such as \GaBiAs) surrounded by a small lattice constant material (such as GaAs), in particular when the relative size of the GaAs region is comparable or larger than the size of the \GaBiAs region. When the diameter of the core region is increased, keeping both Bi fraction and the shell diameter fixed, we find that the strain inside the core region initially becomes more compressive, and then reverses sign towards a tensile strain at large values of $\rm \rho_D$. A \GaBiAs nanowire with $\rm \rho_D$=1.0 is almost entirely under a tensile biaxial strain with the hydrostatic component negligibly small. The \textcolor{blue}{Supplementary Information Figure S2} plots the strain profiles for the small and large core diameter nanowires with D$\rm _S$ =10 nm and 30 nm. The results show a consistent trend in the strain character irrespective of the actual core/shell diameters, confirming that the strain reversal is primarily dependent on the $\rm \rho_D$ parameter. The tuning of the strain from compressive to tensile regime indicates that the \GaBiAs nanowires may provide an excellent platform to manipulate internal strain by simply engineering the core thickness.

\begin{figure*}[t]
\includegraphics[scale=0.245]{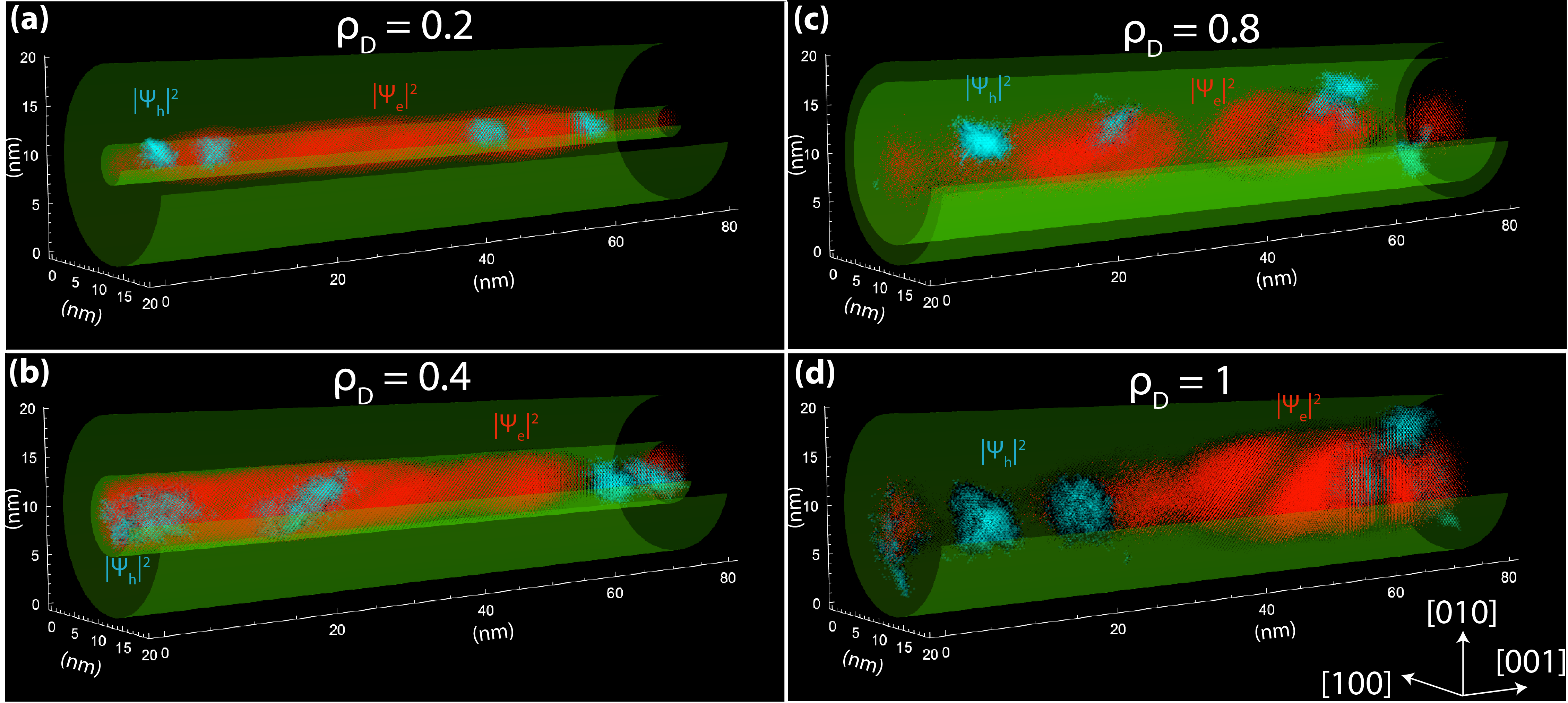}
\caption{\textbf{Wave function confinements.} The three dimensional visualisations of the wave function charge densities are shown for the lowest electron $\rm |\Psi_e|^2$ (shown as red color distribution) and the highest hole $\rm |\Psi_h|^2$ (shown as cyan color distribution) states. The green cylinders are plotted to indicate the boundaries of the core and shell regions of the nanowires. The nanowires are selected with parameters as follows: $x$=15\%, D$\rm _S$=20 nm, $\rm \rho_L$=4 and \textbf{(a)} $\rm \rho_D$=0.2, \textbf{(b)} $\rm \rho_D$=0.4, \textbf{(c)} $\rm \rho_D$=0.8, \textbf{(a)} $\rm \rho_D$=1.0. The plots for other nanowire dimensions are provided in the \textcolor{blue}{Supplementary Information Figure S4}.}
\label{fig:Fig3}
\end{figure*} 

It is well-known in the literature that the highest valence band state is a heavy-hole (HH) type state for a nanostructure under a compressive strain, and conversely the light-hole (LH) state is the ground hole state for a nanostructure under a tensile strain~\cite{Greif_ACSPhotonics_2018, Usman_PRB2_2011, Usman_PRB_2012}. It is also well-established that the HH state which consists of $p_x$ and $p_y$ character dominantly couples to a transverse-electric (TE) mode optical strength, whereas the LH state which has contribution from $p_z$ character also couples to the transverse-magnetic (TM) mode optical strength. For the \GaBiAsGaAs nanowires, we have shown a reversal of strain character from compressive to tensile regime. Therefore, our results suggest that the engineering of $\rm \rho_D$ during the fabrication stage may provide a convenient knob to design nanowire-based optoelectronic devices such as semiconductor optical amplifiers where the polarisation-sensitive light interaction is a desired element. However, we acknowledge that this theoretical prediction is based on the previously established understanding of the relationship between the strain character and the polarisation sensitive optical transitions for the conventional III-V alloys and nanostructures~\cite{T_OP_2005}, and its validation for the \GaBiAsGaAs core$-$shell nanowires still require further experimental advancement.   

In the previous section, the results show a red shift in the absorption wavelength as a function of both nanowire geometry parameter $\rm \rho_D$ and the core Bi fraction $x$. Moreover, the red shift in wavelength as a function of $x$ is stronger for larger core diameter nanowires compared to the narrow core diameters. This dependence of wavelength can be explained by carefully looking at the strain profiles. The strain directly impacts the confinement energies of electron and holes, and therefore plays an important role in the red shift of ground state transition energies~\cite{}. From Figures~\ref{fig:Fig2} (c) and (d), it is clear that the overall magnitude of strain increases when $\rm \rho_D$ is increased. This explains the red shift of wavelength as a function of $\rm \rho_D$ as plotted in Figure~\ref{fig:Fig1}(b). In order to understand the dependence of wavelength on core Bi fraction, we have plotted the strain profiles as a function of $x$ for $\rm \rho_D$=0.2 and 1.0 in the \textcolor{blue}{Supplementary Information Figure S3}. Consistent with Figures~\ref{fig:Fig1} (c) and (d), the hydrostatic and biaxial strain profiles are compressive for $\rm \rho_D$=0.2, and the strain is predominantly tensile biaxial strain for $\rm \rho_D$=1.0. However, we note that the overall magnitude of strain for $\rm \rho_D$=1.0 is much stronger than the magnitude for $\rm \rho_D$=0.2. This explains a stronger red shift in wavelength for $\rm \rho_D$=1.0 when compared to $\rm \rho_D$=0.2 as plotted in Figure~\ref{fig:Fig1}(c).

\noindent
\textbf{\textit{\textcolor{blue}{Spatial confinement of wave functions.}}} Figure~\ref{fig:Fig3} plots the spatial distributions of the lowest electron $\rm \psi_e$ (red color) and the highest hole $\rm \psi_h$ (cyan color) charge densities in the nanowire region for a selected set of $\rm \rho_D$, and with $x$=15\% and D$\rm _S$=20 nm. A complete set of plots are provided in the \textcolor{blue}{Supplementary Information Figure S4}. Notably, we find that for the narrow core nanowires ($\rm \rho_D \leq$0.4), the spatial distributions of the electron wave function are quite uniform throughput the core region when compared to the larger core diameters ($\rm \rho_D \geq$0.8). This can be understood in terms of an interplay between the quantum confinement and strain effects. It is well-known in the literature that the electron wave functions are primarily affected by the character of the hydrostatic strain component~\cite{Usman_IEEE_2009}. The highly compressive hydrostatic strain in conjunction with a strong quantum confinement for the narrow core nanowires lead to a uniformly distributed electron wave function profile. For the large values of $\rm \rho_D$, the hydrostatic strain is significantly relaxed and its impact on the electron confinement is weak. Therefore, the electron confinement in this case will be dominated by the effect of nanowire confinement axis, which is perpendicular to the growth direction. The small or zero thickness of the GaAs shell for $\rm \rho_D \geq$ 0.8 results in a weak quantum confinement effect along the growth direction, and therefore the spatial distributions of the electron wave functions become more localised in the in-plane directions.

The spatial distribution of the hole wave functions is quite different from the electron wave functions and is governed by three effects: quantum confinement, biaxial strain and alloy disorder. As shown in the previous studies for the \GaBiAsGaAs quantum wells~\cite{Usman_APL_2014, Usman_PRA_2018, Usman_PRM_2018}, the impact of the alloy disorder is dominant on the hole wave function confinements which is also evident in our study of nanowires. Overall, we conclude that the strong alloy disorder impact dominates the effects of the quantum confinement and strain, and leads to highly confined hole wave functions as shown in Figure~\ref{fig:Fig3}. 

In the \textcolor{blue}{Supplementary Information Figure S5}, we have plotted the spatial distributions of the electron wave functions for nanowires with the shell diameters of 10 nm and 30 nm at $\rm \rho_D$ = 0.2 and 1.0. Overall, the charge distributions exhibit a very similar character: a stronger confinement of the electron wave functions at $\rm \rho_D$ = 0.2 when compared to $\rm \rho_D$ = 1.0. However, we also note that the effect is much stronger for the shell diameter of 30 nm. This highlight an interesting property of the \GaBiAs nanowires that a stronger electron hole wave function overlap can be achieved for the narrow core nanowires with small shell diameters. On the other hand, the fabrication of nanowires with large shell diameters accompanied with large values of $\rm \rho_D$ is expected to result in a stronger confinement of the electrons and therefore a larger separation of the electron and hole charge carriers. The enhanced separation of the charge carriers could potentially be useful for applications requiring low carrier recombination rates such as light harvesting in solar cells. The larger charge separation in the \GaBiAs nanowires with $\rm \rho_D \geq$ 0.8 may lead to a higher carrier collection at the different ends of the nanowire region and hence improve the efficiency of the solar to electrical conversion. Given that there is currently a significant research interest in employing bismide materials to target 1 eV spectral range in the design of multi-junction photovoltaic devices~\cite{Richards_SEMSC_2017, Kim_JCG_2016, Usman_PRA_2018}, this insight could be a catalysis for the experimental efforts to explore \GaBiAs nanowires in the future solar cell devices. 

\begin{figure}[t]
\includegraphics[scale=0.35]{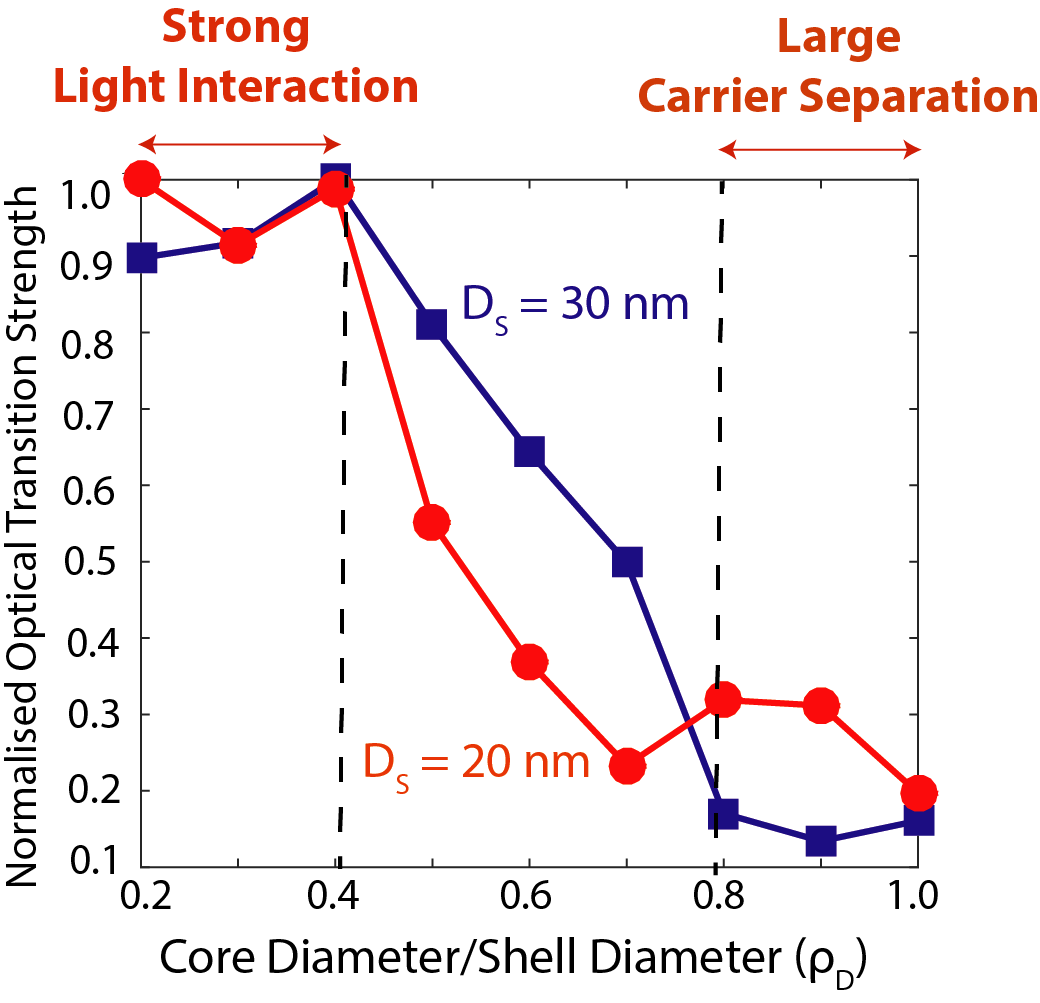}
\caption{\textbf{Optical absorption strength.} The plots of the normalised ground-state inter-band optical absorption strengths are shown as a function of $\rm \rho_D$ for $x$=15\%, $\rm D_S$=20 nm and 30 nm. In both nanowire structures, the optical absorption strength is significantly stronger at small diameters of the \GaBiAs core regions ($\rm \rho_D \leq$ 0.4). The stronger confinement of both electron and hole states at large core diameters ($\rm \rho_D \geq$ 0.8) offers higher charge carrier separation which could potentially be useful for solar to electric conversion applications.}
\label{fig:Fig4}
\end{figure}

\noindent
\textbf{\textit{\textcolor{blue}{Inter-band absorption strength.}}} The spatial distributions of the electron and hole wave functions directly impact the strength of the inter-band optical transitions, which is proportional to the electron/hole wave function overlaps. In the previous section, we have shown that the spatial profiles of the electron and hole wave functions are strongly dependent on the geometry parameters of the nanowires, therefore we expect a commensurate dependence for the optical transition strengths. Figure~\ref{fig:Fig4} plots the normalised inter-band optical transition strengths as a function of $\rm \rho_D$ for Bi=15\%, D$\rm _S$=20 nm and 30 nm. For both geometry parameters of the nanowires, we compute that the transition strengths are significantly higher for $\rm \rho_D \leq$ 0.4 compared to $\rm \rho_D \geq$ 0.8. This is directly related to the electron wave function spatial distributions which are uniform in the core region for the nanowires with $\rm \rho_D \leq$ 0.4, leading to a higher overlap with the strongly confined hole wave functions. In the discussion above, we have also shown that for $\rm \rho_D \leq$ 0.4, the inhomogeneous broadening due to the random alloy configurations is significantly less compared to $\rm \rho_D \geq$ 0.9, therefore we propose that the future fabrication of nanowires for telecom-wavelength photonic devices such as lasers, photodetectors, and sensors could target narrow core nanowires with relatively high Bi fractions (around 30\%). Finally, we note that our calculations of interband optical absorption strengths exhibit consistent trend for the two different shell diameters (20 nm and 30 nm). Therefore, we conclude that the tuning of the interband absorption strength only require to engineer $\rm \rho_D$ and does not impose restrictions on the actual core/shell diameters and the length of the nanowire.   

\begin{figure*}[t]
\includegraphics[scale=0.26]{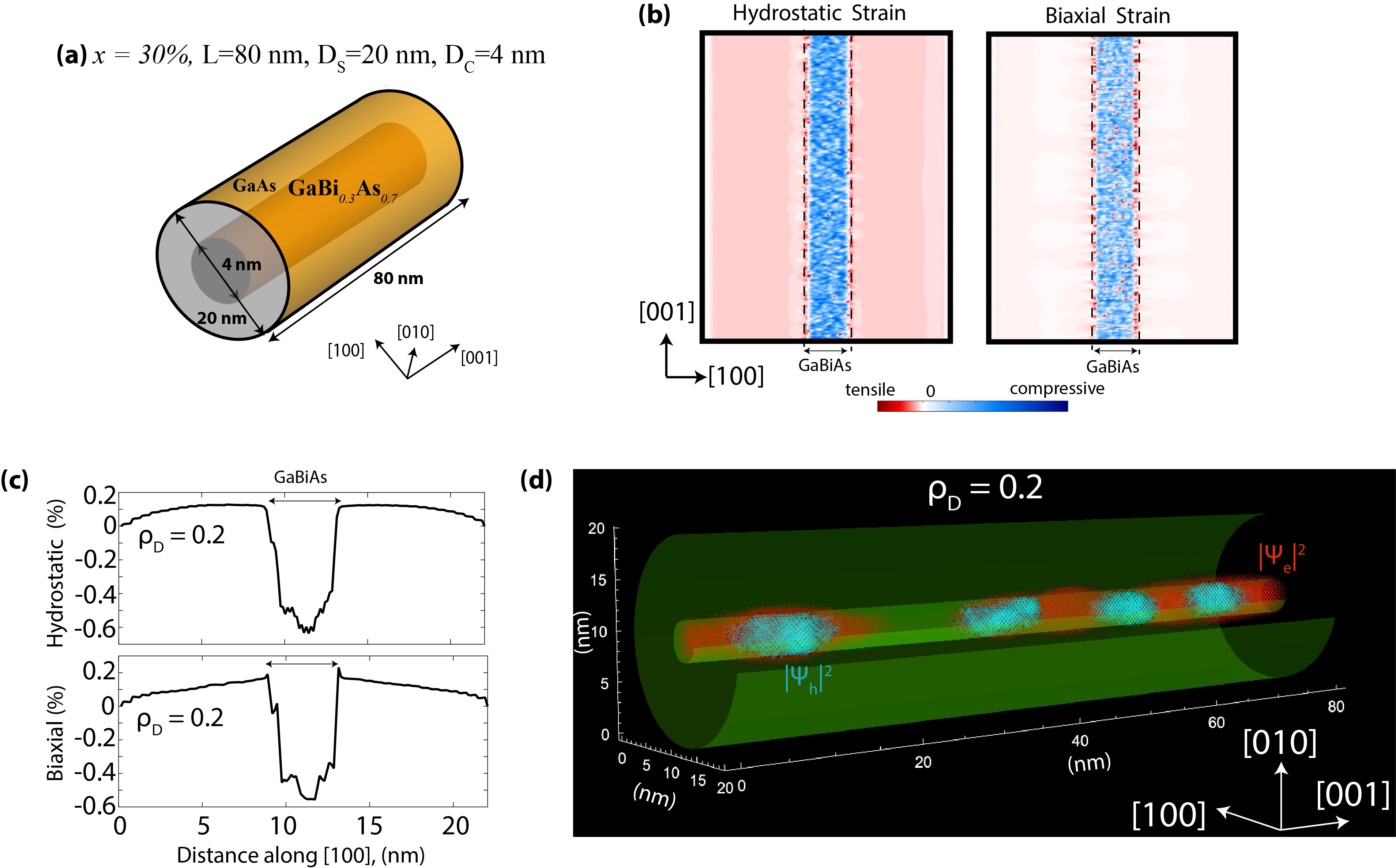}
\caption{\textbf{An exemplary nanowire with optimised geometry.} \textbf{(a)} The designed geometry of a nanowire for optimal optical absorption strength at telecom-wavelength is illustrated. The core$-$shell nanowire has a core diameter of 4 nm and a shell diameter of 20 nm with 30\% Bi fraction in the central \GaBiAs core region. \textbf{(b)} The plots of the hydrostatic and biaxial strain components are shown for the nanowire structure sketched in \textbf{(a)}. The color plots show the strain profiles in a 2D [010]-plane through the center of the nanowires. The blue (red) color regions indicate the presence of a compressive (tensile) strain. \textbf{(c)} The 1D plots of the hydrostatic and biaxial strain components are shown through the center of the nanowire. The computed strain at each data point along the [100] direction represents an average over all data points in the corresponding [100] plane and for five different random configurations of Bi atoms. The nanowire core region is compressively strained, whereas the strain in the shell region is quite small. \textbf{(d)} The three-dimensional visualisations of wave function charge densities are shown for the lowest electron (shown as red color distribution) and the highest hole (shown as cyan color distribution) states. The green cylinders are plotted to indicate the boundaries of the core and shell regions.}
\label{fig:Fig5}
\end{figure*} 

\noindent
\textbf{\textit{\textcolor{blue}{Optimised nanowires for $\rm 1.55 \, \upmu \rm m$ devices.}}} In the preceding discussions, our large-scale atomistic simulations have shown that nanowires with narrow core diameters ($\rm \rho_D \leq$ 0.4) lead to a relatively strong inter-band optical transition strength at telecom-wavelength (1.55 $\upmu$m). To confirm this prediction and provide directly relevant results, we simulated an exemplary nanowire with $\rm \rho_D$=0.2, L=80 nm and $x$=30\% as shown in Figure~\ref{fig:Fig5} (a). The ground state inter-band transition wavelength is computed at 1.55 $\upmu$m with an inhomogeneous broadening of $\approx$ 30 nm, which is consistent with the expectation based on Figure~\ref{fig:Fig1} for $\rm \rho_D$=0.2. The strain profiles for this nanowire are shown in Figure~\ref{fig:Fig5} (b) and (c), which again exhibit quite similar trends as computed for $x$ = 15\%. The strain is highly compressive inside the core region and its magnitude is small in the shell region. The 1D cuts plotted in (c) further confirm this character of the strain. Interestingly, we note that the overall magnitude of the strain is about a factor of two larger than its magnitude at $x$=15\%. This is expected because for the same volume of the core region, doubling the Bi fraction (from 15\% to 30\%) implies a relatively commensurate increase in the lattice-mismatch between GaAs and \GaBiAs, which will correspondingly increase the magnitude of the strain in the system. The plots of the lowest electron and the highest hole wave function distributions are shown in (d) by using red and cyan colors, respectively. The electron wave function is spread inside the large parts of the core region which leads to a strong inter-band optical absorption strength. The computed inter-band optical absorption strength is found to be consistent with the values shown earlier for $\rm \rho_D$=0.2 and $x$=15\%. These results reinforce our conclusion that the \GaBiAsGaAs core$-$shell nanowires designed with around 30\% Bi fraction in the core region and $\rm \rho_D \leq$ 0.4 are expected to provide favourable properties for photonic devices working at telecom-wavelengths.  

In terms of Bi incorporation in the GaAs material system, there has been significant experimental progress over the last decade~\cite{Bismuth_containing_compounds_2013, Bismuth_containing_alloys_2019}. Recent studies have shown up to 20\% Bi in GaAs with good quality optoelectronic characteristics, and further studies are in-progress to increase the Bi contents up to 30\%~\cite{Bismuth_containing_alloys_2019}. For core$-$shell nanowires, the recent study has reported a Bi incorporation of 10$\pm$2\%~\cite{Oliva_arXiv_2019}, which is significantly higher than the $\sim$2\% Bi reported in the first study~\cite{Ishikawa_Nanoletters_2015}. The incorporation of about 30\% Bi in nanowires for the proposed optimised light absorption will require further experimental progress in the coming years, however, it is anticipated that the near term devices may be based on relatively low Bi fractions (5-15\%) accompanied by larger $\rm \rho_D$, which could be suitable for applications where a large carrier separation is desirable.      

Finally, we wish to point out that the spatial confinements of the highest hole wave functions in the investigated nanowires are in stark contrast with the quantum well structures previously studied in the literature~\cite{Usman_JPCM_2019}. At the large Bi fractions ($x \geq$10\%), the impact of the random alloy configurations was found to becomes weak and the confinement of the hole wave functions became quite uniform in the quantum well nanostructures. However, this is not applicable in the case of the studied nanowires, where the results suggest that even at $x$=30\%, the hole wave functions are strongly confined in localised regions of the nanowire, indicating a relatively stronger impact of the random alloy configurations on the hole wave function profile.

We note that this work is focused on the \GaBiAs material, however other bismide materials such as GaP$_{1-x}$Bi$_x$, In$_x$Ga$_{1-x}$Bi$_y$As$_{1-y}$, and GaBi$_x$N$_y$As$_{1-x-y}$ have also exhibited similar promising band structure properties based on the published bulk and quantum well studies~\cite{Bushell_SR_2019, Broderick_PSSB_2013, Usman_PRA_2018, Bismuth_containing_compounds_2013, Bismuth_containing_alloys_2019}. Therefore, it would be interesting to investigate and compare the optoelectronic properties of core$-$shell nanowires for a range of bismide materials, which will be a topic of a future study in conjunction with experimental efforts in this direction.

\section{Conclusions} In conclusion, we have presented a detailed understanding of the electronic and optical properties of the \GaBiAsGaAs core$-$shell nanowires by performing million-atom atomistic simulations. For the technologically relevant telecom-wavelength photonic devices, our results predicted two pathways to design nanowires: either by increasing Bi fraction with a small core/shell diameter ratio ($\rm \rho_D$), or by increasing $\rm \rho_D$ at around 15\% Bi fraction. A comparative analysis of the two options revealed that a smaller inhomogeneous broadening and a stronger light interaction can be achieved by designing nanowires with $\rm \rho_D \leq$ 0.4. This highlighted a favourable direction for the future experiments on the fabrication of the \GaBiAsGaAs nanowires. Our results also suggested that nanowires with $\rm \rho_D \geq$ 0.8 may lead to a higher separation between the electron and hole charge carriers, which could be useful for other applications demanding low recombination rates such as photovoltaic devices. The computed reversal of the strain character from a compressive to a tensile regime unveiled an interesting property of the investigated \GaBiAsGaAs nanowires, which could be exploited for optoelectronic devices where a polarisation sensitive light emission/absorption is required. The presented results, based on the well-benchmarked theoretical models, provide a highly reliable quantitative guidance to the ongoing experiments in an emerging area of research and may strongly contribute in advancement of the future nanoelectronic and nanophotonic technologies.        

\noindent
\section{Computational Methods}
A detailed description of the computational methods is provided in the accompanied \textcolor{blue}{supplementary information document} and a summary of the methodologies is as follows. The atomic structures of the nanowire system are relaxed using the valence force field model (VFF)~\cite{Keating_PR_1966}. For the VFF model, we have used the experimental bulk lattice constants and elastic coefficients~\cite{Usman_PRB_2011}. After the atomic relaxation, the strain of the atoms is computed, and the results for different strain components are shown and discussed in Figure~\ref{fig:Fig2}. The electronic structure is computed based on ten-band sp$^3$s$^*$ tight-binding parameters including spin-orbit coupling~\cite{Usman_PRB_2011}. The diagonalization of the tight-binding Hamiltonian at $\Gamma$ point provides the electron and hole energies and wave functions. The ground state optical transition wavelength is computed from the lowest electron and the highest hole energies. The atomistic charge densities for the corresponding wave functions are visualised in real-space over the whole nanowire region by plotting red and cyan dots at the location of each atom, with the intensity of the color modulated by the strength of the charge density at that atomic sight (Figure~\ref{fig:Fig3} and ~\ref{fig:Fig5} (c)). The inter-band optical transition strengths plotted in Figure~\ref{fig:Fig4} are computed by using the Fermi's golden rule~\cite{Usman_APL_2014}.   

\noindent
\\
\textbf{Acknowledgements:} Computational resources are acknowledged from the National Computational Infrastructure (NCI) under the National Computational Merit based Allocation Scheme (NCMAS). The author thanks Viktor Perunicic for useful discussions about the visualisation of the wave function charge densities. Some aspects of this work were done as part of the EU FP7 Project BIANCHO. 
\\ \\
\noindent
\textbf{Data availability:}
The data that support the findings of this study are available within the article and its supplementary information file. Further requests can be made to the corresponding author. 
\\ \\
\noindent
\textcolor{black}{
\textbf{Additional information:}} A supporting information file consists of additional details on methods and extended data on strain and wave function profiles. 
\\ \\
\noindent
\textbf{Competing financial interests:} The author declare no competing interests. 

\clearpage
\newpage

\begin{center}
\Large{\textbf{\underline{Supplementary Information Document}}}
\end{center}

\normalsize

\noindent
\section{Computational Methods}

\noindent
\textbf{Geometry Parameters:}
The schematic diagram of the investigated \GaBiAsGaAs nanowire is shown in the Figure 1(a) of the main text. In our study, the core$-$shell nanowires consist of a \GaBiAs core region with diameter D$\rm _C$, length L, and Bi fraction $x$. The shell region is made up of GaAs material with diameter D$\rm _S$ and length L. The largest nanowire with D$\rm _S$=30 nm and L=80 nm consists of about 3.5 million atoms and the smallest nanowire with D$\rm _S$=10 nm and L=80 nm consists of about 0.37 million atoms. The simulations over large structures with atomistic resolution and realistic boundary conditions allow us to provide a highly reliable theoretical analysis of the nanowire electronic and optical properties.    

In our simulations, the Bi atoms are randomly placed replacing the As atoms in the core region of the nanowires. The random spatial distribution of the Bi atoms is dependent on a four-digit seed value input to a random number generator that determines the nature of an anion atom (either Bi or As) at a given atomic location inside the nanowire core region. Different seed values ensure different spatial arrangements of the Bi atoms, resulting in statistically different numbers and types of Bi pairs and clusters in the core region. All of the results presented in this study are based on an average over five different random distributions of the Bi atoms. 

It is critically important to simulate a large size for the \GaBiAs supercell to properly model the alloy disorder effects ~\cite{Usman_PRB_2013}. In our previous study on quantum well structures, We have probed the electron and hole energies as well as the associated inhomogeneous broadening in the ground state transition energies when the strained \GaBiAs supercell size is increased from 1000 atoms to 512000 atoms~\cite{Usman_PRM_2018}. Our calculations showed that the small size of supercells ($<$ 4096 atoms) artificially modifies the electron/hole energies and enhance the strength of the inhomogeneous broadening. By increasing the supercell size from 8000 atoms to 512000 atoms, we found that the electron and hole energies changed by less than 1 meV and 10 meV respectively, whereas the values of the inhomogeneous broadening for the hole energies were roughly 27 meV in good agreement with the measured value of 31 meV~\cite{Usman_PRB_2013}. We therefore believe that a supercell size consisting of 8000 atoms or more is suitable to provide a reliable estimate of the properties of the strained \GaBiAs alloys. In this work on nanowires, we have performed large-scale atomistic simulations consisting of up to 3.5 million atoms in the over supercell, including the number of atoms in the \GaBiAs core region above 50000 in all cases. 

\noindent
\textbf{Calculation of Strain:}
In order to calculate the strain induced by the lattice-mismatch between the GaAs and \GaBiAs materials, the \GaBiAsGaAs nanowires are relaxed by applying atomistic valence force field (VFF) energy minimization scheme~\cite{Keating_PR_1966, Lazarenkova_APL_2004, Usman_PRB_2011}. The VFF parameters for the GaBi and GaAs materials are published in the literature~\cite{Usman_PRB_2011}. The values of $\alpha_{0}$ and $\beta_{0}$ for the GaAs are taken from Lazarenkova \textit{et al}.~\cite{Lazarenkova_APL_2004}, whereas for GaBi are determined by obtaining relaxed bond lengths in accordance with Kent \textit{et al.}~\cite{Kent_PRB_2001}. 

After the VFF relaxation, the relaxed Ga-Bi bond-lengths are found to follow the trends of $x$-ray absorption spectroscopy measurements~\cite{Usman_PRB_2011}. The strain tensor components ($\epsilon_{xx}, \epsilon_{yy}, \epsilon_{zz}, \epsilon_{xy}, \epsilon_{yz}, \epsilon_{xz}$) are computed from the relaxed bond-lengths of atoms~\cite{Lazarenkova_APL_2004}. The strain parameters of interest, which are directly related to the energy shifts in the conduction and valence band edges are hydrostatic ($\epsilon_H$) and biaxial ($\epsilon_B$) strain components, which are defined as follows~\cite{Usman_PRB2_2011}: 
\noindent
\begin{eqnarray}
\epsilon_{\textrm H} &=& \epsilon_{xx} + \epsilon_{yy} + \epsilon_{zz} \\
\epsilon_{\textrm B} &=& 2 \epsilon_{zz} - \epsilon_{xx} - \epsilon_{yy}
\label{eq:strains}
\end{eqnarray}

\noindent
\textbf{Electronic and Optical Simulations:}
The electronic structure of the studied nanowires is computed from the nearest-neighbour ten-band \spss tight-binding (TB) theory, which explicitly include spin-orbit coupling. The TB parameters for the GaAs and GaBi materials were published in our previous study~\cite{Usman_PRB_2011}, which accurately reproduced the bulk band structure of these two materials computed from DFT model~\cite{Janotti_PRB_2002}. In the published studies on both unstrained and strained GaBi$_{x}$As$_{1-x}$ alloys, the tight-binding model has shown good agreement with a series of available experimental data sets~\cite{Donmez_SST_2015, Balanta_JoL_2017, Zhang_JAP_2018, Dybala_APL_2017, Collar_AIPA_2017, Usman_PRB_2011, Usman_PRB_2013, Broderick_PRB_2014}, as well as with the DFT calculations reported in the literature~\cite{Kudrawiec_JAP_2014, Polak_SST_2015, Bannow_PRB_2016}.

By solving the tight-binding Hamiltonian, we obtain ground state electron and hole energies and the corresponding wave functions at the $\Gamma$ point ($k$=0), which are labelled as $\vert \psi_{e} \rangle$ and $\vert \psi_{h} \rangle$, respectively, and are defined as: 
\noindent
\begin{eqnarray}
\vert \psi_{e} \rangle &=& \sum_{i,\mu} C^e_{i,\mu} \vert i\mu \rangle  \\
\vert \psi_{h} \rangle &=& \sum_{j,\nu} C^h_{j,\nu} \vert j\nu \rangle
\label{eq:e_h_wfs}
\end{eqnarray}
\noindent
where the label $i$ ($j$) represents the atom number inside the supercell and $\mu$ ($\nu$) denotes the orbital basis states on an atom for the electron (hole) states. $C^e_{i,\mu}$ and $C^h_{j,\nu}$ are the coefficients of the electron and hole wave functions, respectively, computed by diagonalising the TB Hamiltonian. 

The overlap between the electron and hole ground states is computed as follows:
\noindent
\begin{eqnarray}
\vert \langle \psi_{h} \vert \psi_{e} \rangle \vert &=& \vert  \sum_{i = j} \sum_{\mu = \nu} (C^h_{j,\nu})^* C^e_{i,\mu} \vert
\label{eq:e_h_overlap}
\end{eqnarray}
\noindent

The inter-band optical transition strength between the ground electron and hole states is computed by first computing the momentum matrix element as follows~\cite{Usman_PRB2_2011}:
\noindent
\begin{eqnarray}
M_{\overrightarrow{n}}^{\alpha \beta} &=& \sum_{i,j} \sum_{\mu,\nu} (C^e_{i,\mu,\alpha})^* (C^h_{j,\nu,\beta}) {\langle i\mu\alpha \vert \textrm{\textbf{H}} \vert j\nu\beta \rangle} {(\overrightarrow{n}_{i}-\overrightarrow{n}_{j})} \label{eq:momentum_x}
\end{eqnarray}
\noindent
where $\alpha$ and $\beta$ represent spin of states, \textbf{H} is the \spss tight-binding Hamiltonian, and $\overrightarrow{n} = \overrightarrow{n}_{i} - \overrightarrow{n}_{j}$ is the real space displacement vector between atoms $i$  and $j$, and is equal to  $\overrightarrow{x}_{i} - \overrightarrow{x}_{j}$. The optical transition strengths is then calculated by using the Fermi's Golden rule and summing the absolute values of the momentum matrix elements over the spin degenerate states:
\noindent
\begin{eqnarray}
\textrm{Optical Absorption} &=&  \sum_{\alpha, \beta} \vert M_{\overrightarrow{x}}^{\alpha \beta} \vert ^2 \label{eq:TE_X}
\end{eqnarray}   

The tight-binding model is implemented with in the framework of atomistic tool NanoElectronic Modeling (NEMO 3-D) simulator~\cite{Klimeck_IEEETED_2007_1} which has, in the past, shown an unprecedented accuracy to match experiments for the study of nano-materials~\cite{Usman_PRB_2011, Usman_PRB_2013, Usman_NN_2016} and devices~\cite{Usman_PRB2_2011, Usman_IOP_2012}. 

\renewcommand{\thefigure}{\textbf{S\arabic{figure}}}
\renewcommand{\figurename}{\textbf{Supplementary Fig.}}

\renewcommand{\thetable}{\textbf{T\arabic{table}}}
\renewcommand{\tablename}{\textbf{Supplementary table.}}

\setcounter{figure}{0}

\begin{figure*}[t]
\includegraphics[scale=0.4]{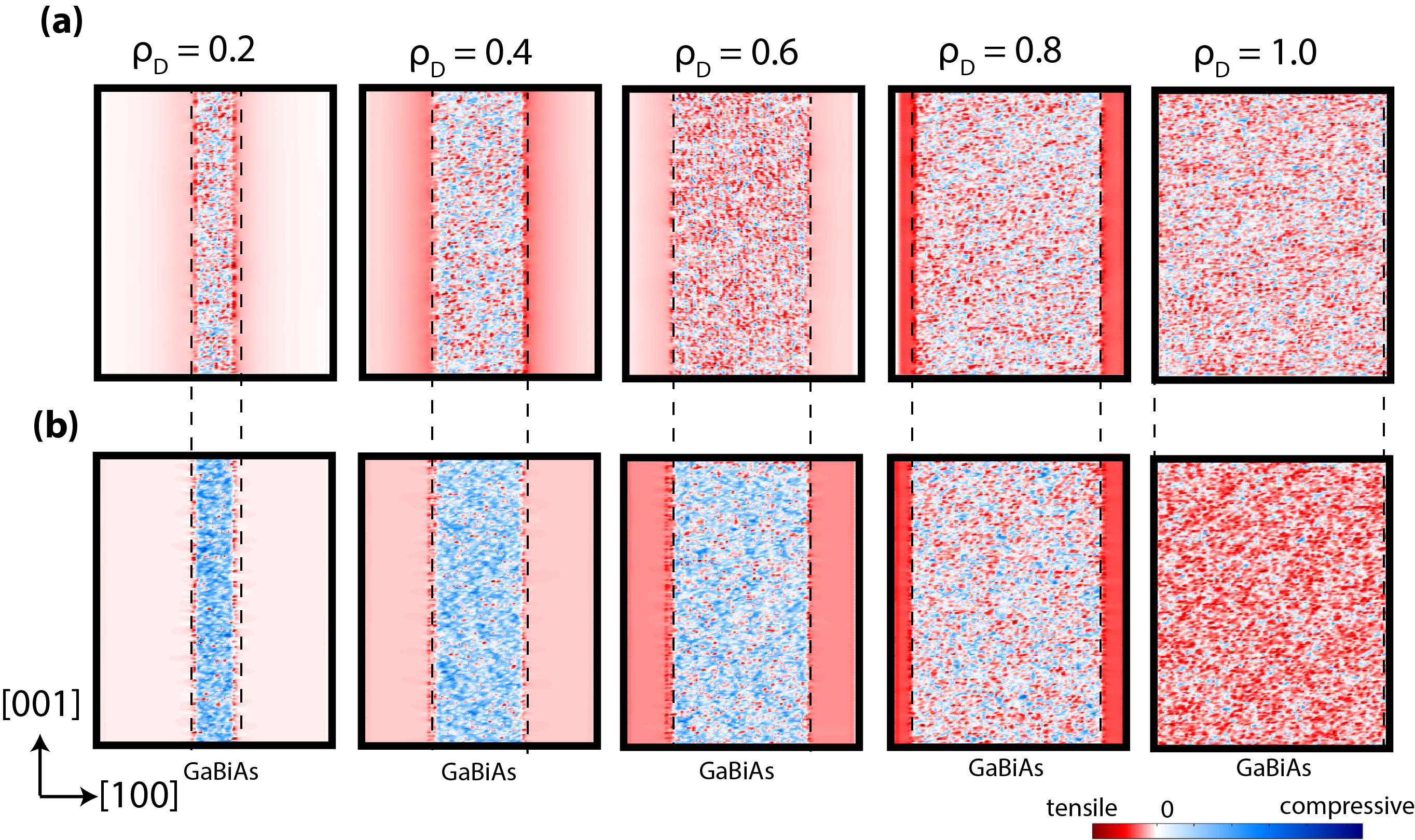}
\caption{\textbf{(a)} The plots of strain tensor component ($\epsilon_{xx}$) are shown for the \GaBiAsGaAs nanowire with $x$=15\%, $\rm \rho_L$=4, a few selected values of $\rm \rho_D$. The color plots show the strain profile in a 2D plane through the center of the nanowire. The blue (red) color regions indicate the presence of compressive (tensile) strain. \textbf{(b)} The same as \textbf{(a)} but for the strain tensor component ($\epsilon_{zz}$).}
\label{fig:Fig1s}
\end{figure*}

\begin{figure*}[t]
\includegraphics[scale=0.35]{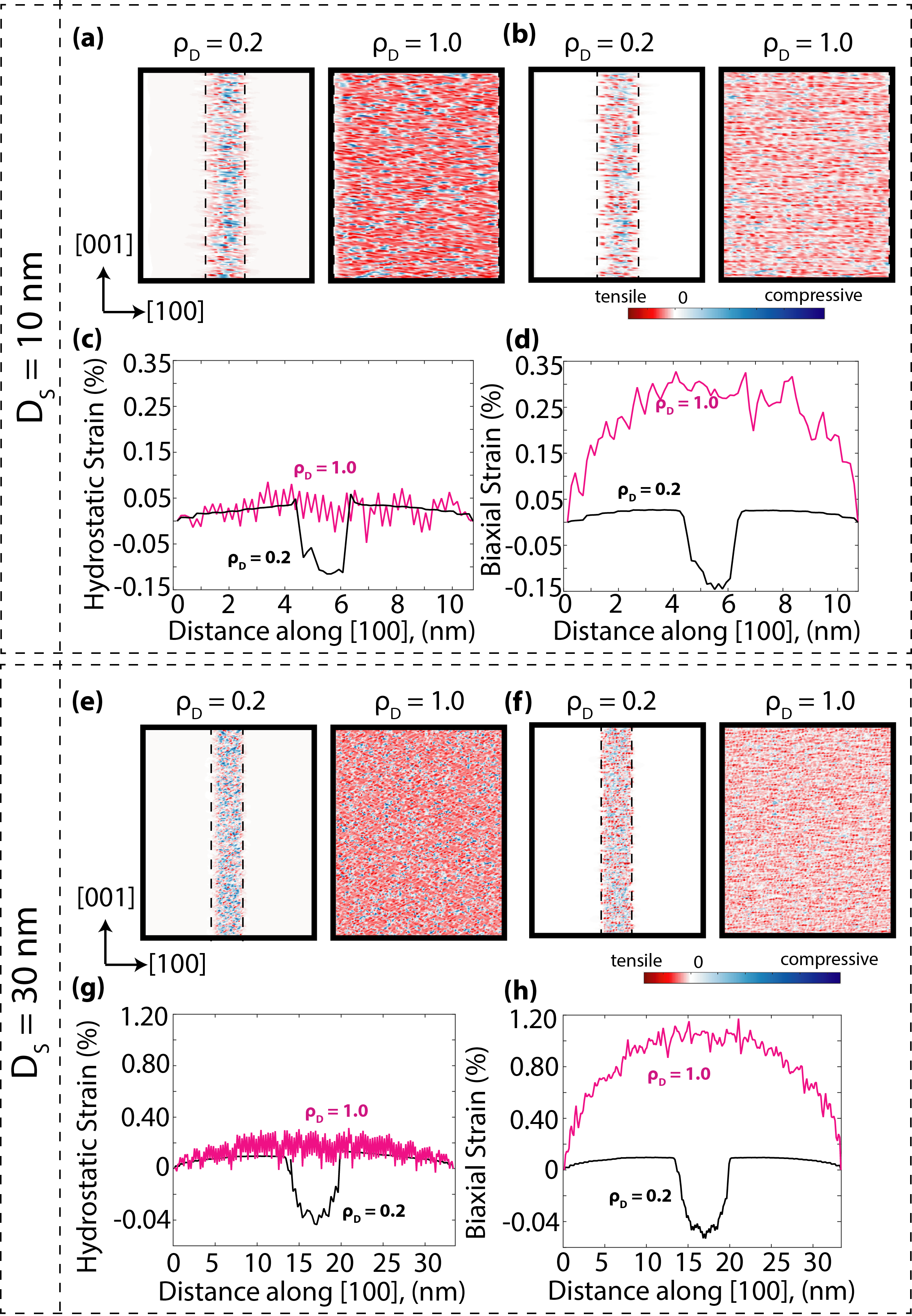}
\caption{\textbf{(a,b)} The plots of the hydrostatic and biaxial strain components are shown for the nanowire structure with $x$=15\%, D$\rm _S$=10 nm, $\rm \rho_L$=8, a couple of selected values of $\rm \rho_D$. The color plots show the strain profiles in a 2D [010]-plane through the center of the nanowires. The blue (red) color regions indicate the presence of a compressive (tensile) strain. \textbf{(c,d)} The 1D plots of the hydrostatic and biaxial strain components are shown through the center of the nanowire. The computed strain at each data point along the [100] direction represents an average over all data points in the corresponding [100] plane and for five different random configurations of Bi atoms. \textbf{(e-h)} The same as \textbf{(a-d)} but for the nanowire structures with $x$=15\%, D$\rm _S$=30 nm, $\rm \rho_L$=2.67, a couple of selected values of $\rm \rho_D$.}
\label{fig:Fig2s}
\end{figure*}

\begin{figure*}[t]
\includegraphics[scale=0.4]{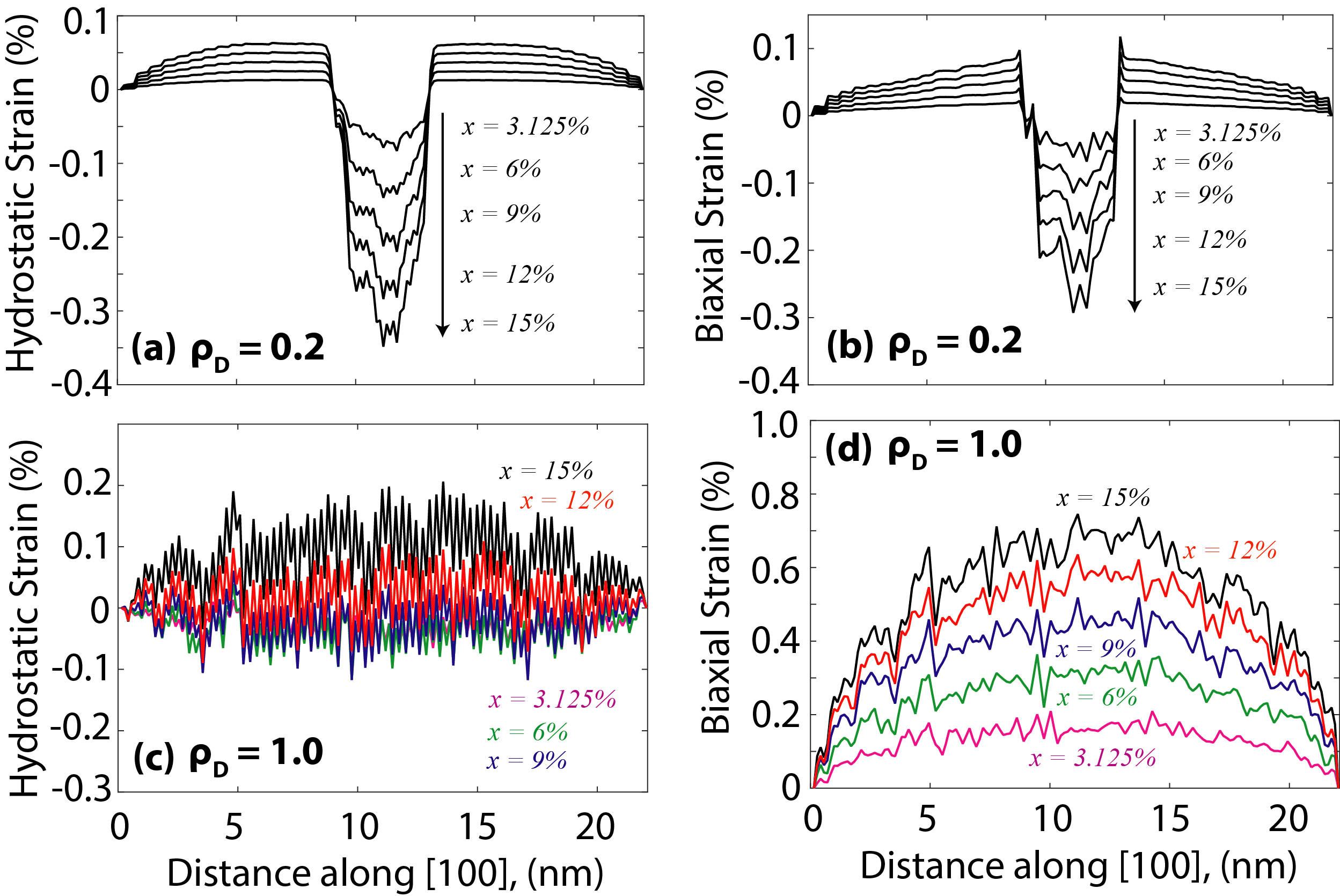}
\caption{\textbf{(a,b)} The 1D plots of the hydrostatic and biaxial strain components are shown as a function of the Bi fraction ($x$) in the core region  through the center of the nanowire with $\rm \rho_D$=0.2. The computed strain at each data point along the [100] direction represents an average over all data points in the corresponding [100] plane and for five different random configurations of Bi atoms. \textbf{(c,d)} The same as \textbf{(a,b)} but for the nanowire structures with $\rm \rho_D$=1.0.}
\label{fig:Fig3s}
\end{figure*}

\begin{figure*}[t]===
\includegraphics[scale=0.245]{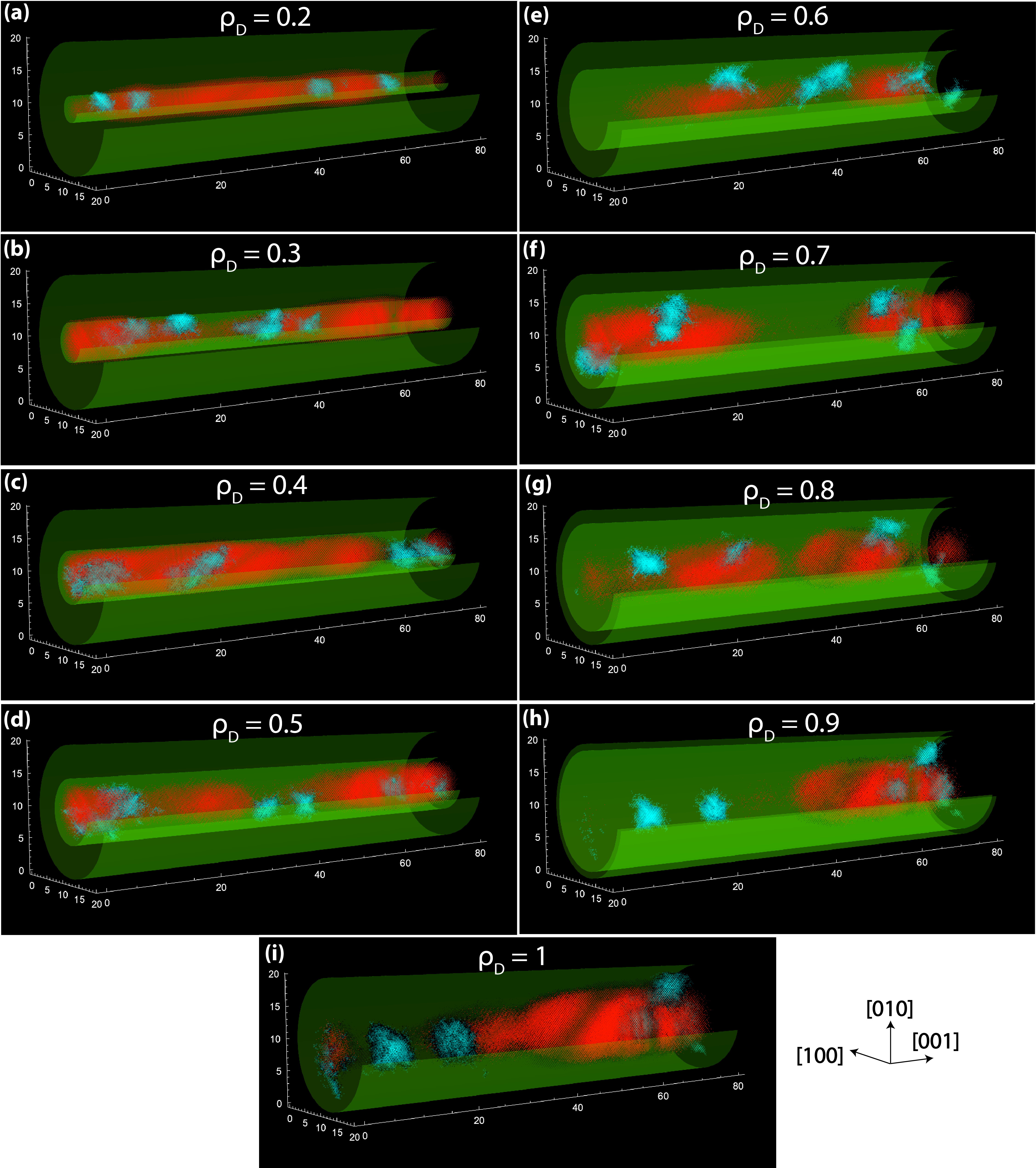}
\caption{The three dimensional visualisations of charge densities are shown for the lowest electron (shown as red color distribution) and the highest hole (shown as cyan color distribution) states. The green cylinders are plotted to indicate the boundaries of the core and shell regions. The nanowires are selected with parameters as follows: D$\rm _S$=20 nm, $x$=15\%, $\rm \rho_L$=4 and the values of $\rm \rho_D$ as marked on the plots.}
\label{fig:Fig4s}
\end{figure*}

\begin{figure*}[t]
\includegraphics[scale=0.22]{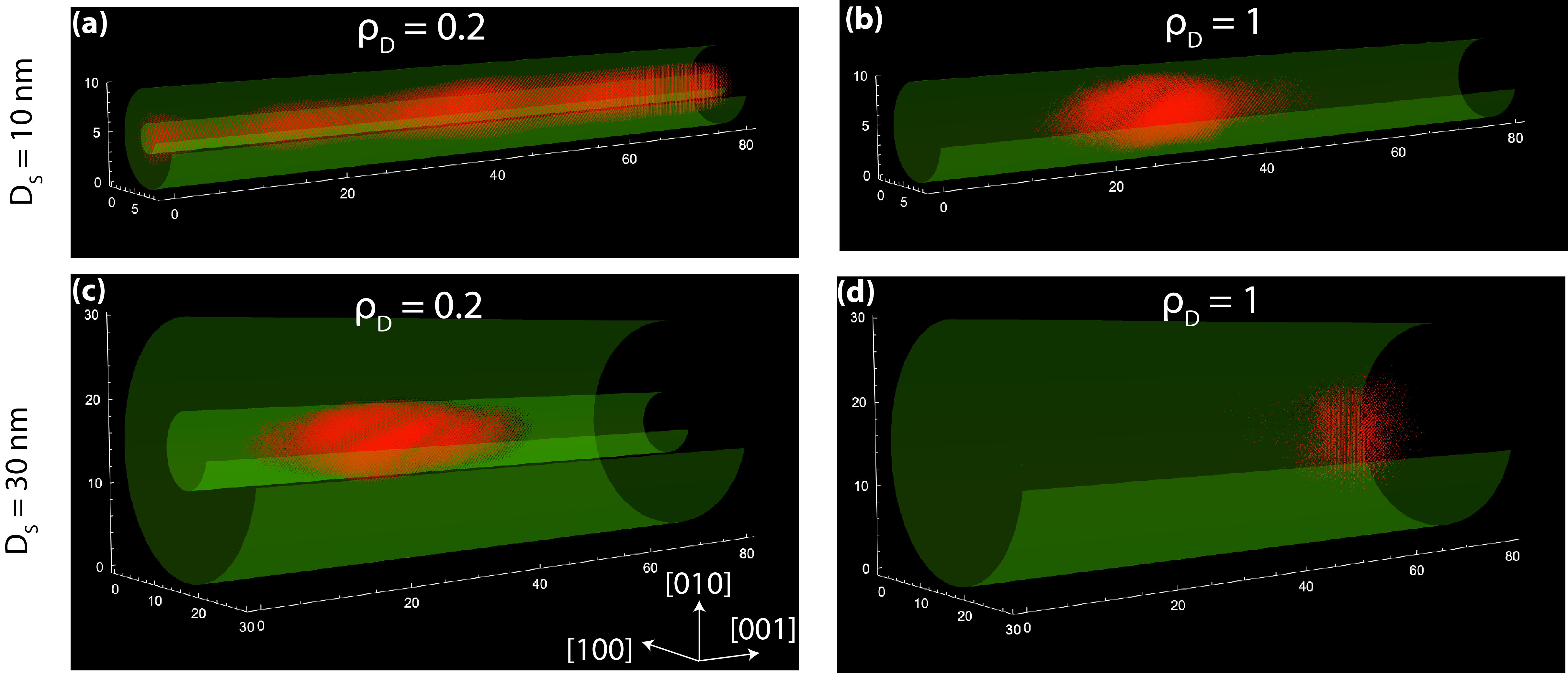}
\caption{The three dimensional visualisations of charge densities are shown for the lowest electron (shown as red color distribution) and the highest hole (shown as cyan color distribution) states. The green cylinders are plotted to indicate the boundaries of the core and shell regions. The nanowires in (a,b) are selected with parameters as follows: D$\rm _S$=10 nm, $x$=15\% and $\rm \rho_L$=4. The nanowires in (c,d) are selected with parameters as follows: D$\rm _S$=30 nm, $x$=15\% and $\rm \rho_L$=4. The selected values of $\rm \rho_D$ are marked on the plots.}
\label{fig:Fig5s}
\end{figure*}

\clearpage
\newpage

%\bibliography{GaNAsBi_electronic_structure}
\def\bibsection{\subsection*{\refname}}

\bibliographystyle{naturemag}
%\bibliography{GaNAsBi_electronic_structure}

\end{document}